\documentclass[journal]{IEEEtran}

\usepackage{graphics,colortbl,multicol,lipsum,xcolor,graphicx,color}
\usepackage[cmex10]{amsmath}
\usepackage{amsthm,mathrsfs,mathtools,amsbsy,amsfonts,amssymb}
\usepackage{setspace,float,pstool,lettrine,newclude}
\usepackage[normalem]{ulem}
\usepackage{latexsym,algpseudocode,gensymb,balance,multirow,bm,wasysym}
\usepackage{cite}
\usepackage[linesnumbered,ruled,vlined]{algorithm2e}
\SetKwInput{KwInput}{Input}  
\SetKwInput{KwOutput}{Output}
\SetKw{KwBy}{by}
\makeatletter
\newcommand{\nosemic}{\renewcommand{\@endalgocfline}{\relax}}
\newcommand{\dosemic}{\renewcommand{\@endalgocfline}{\algocf@endline}}
\let\oldnl\nl
\newcommand{\nonl}{\renewcommand{\nl}{\let\nl\oldnl}}
\makeatother
\ifCLASSOPTIONcompsoc
\usepackage[caption=false,font=normalsize,labelfon
t=sf,textfont=sf]{subfig}
\else
\usepackage[caption=false,font=footnotesize]{subfig}
\fi

\DeclarePairedDelimiterX\MeijerM[3]{\lparen\!}{\rparen}%
{\,#3\delimsize\vert\begin{smallmatrix}#1 \\ #2\end{smallmatrix}}

\newcommand\MeijerG[8][]{%
  G^{\,#2,#3}_{#4,#5}\MeijerM[#1]{#6}{#7}{#8}}

\WithSuffix\newcommand\MeijerG*[7]{%
  G^{\,#1,#2}_{#3,#4}\MeijerM*{#5}{#6}{#7}}

\allowdisplaybreaks

\newtheorem{proposition}{Proposition}

\graphicspath{{figures/}}
\allowdisplaybreaks

\newcommand{\RNum}[1]{\uppercase\expandafter{\romannumeral #1\relax}}

\usepackage{mathtools}

\usepackage{accents}
\newlength{\dhatheight}
\newlength{\dtildeheight}

\begin{document}

\title{Semantics-Aware Source Coding \\in Status Update Systems}
\author{
    \IEEEauthorblockN{Pouya Agheli\IEEEauthorrefmark{1}, Nikolaos Pappas\IEEEauthorrefmark{2}, and Marios Kountouris\IEEEauthorrefmark{1}}\\
    \IEEEauthorblockA{\IEEEauthorrefmark{1}EURECOM, Communication Systems Department, Sophia Antipolis, France}\\
\IEEEauthorblockA{\IEEEauthorrefmark{2}Dept. of Science and Technology, Link\"{o}ping University, Norrk\"{o}ping Campus, Sweden\\ Email: pouya.agheli@eurecom.fr, nikolaos.pappas@liu.se, marios.kountouris@eurecom.fr}
\thanks{The work of P. Agheli and M. Kountouris has received funding from the European Research Council (ERC) under the European Union’s Horizon 2020 research and innovation programme (Grant agreement No. 101003431). The work of N. Pappas is supported by the Swedish Research Council (VR), ELLIIT, and CENIIT.}
}
\maketitle

\begin{abstract}
We consider a communication system in which the destination receives status updates from an information source that observes a physical process. The transmitter performs semantics-empowered filtering as a means to send only the most ``important" samples to the receiver in a timely manner. 
As a first step, we explore a simple policy where the transmitter selects to encode only a fraction of the least frequent realizations of the observed random phenomenon, treating the remaining ones as not informative. 
For this timely source coding problem, we derive the optimal codeword lengths in the sense of maximizing a semantics-aware utility function and minimizing a quadratic average length cost. Our numerical results show the optimal number of updates to transmit for different arrival rates and encoding costs and corroborate that semantic filtering results in higher performance in terms of timely delivery of important updates. 
\end{abstract}

\IEEEpeerreviewmaketitle

\section{Introduction}
The evolution of the latest generations of mobile communication systems has been mainly driven by setting highly ambitious, often hard to achieve, goals. Although this maximalistic approach may trigger technological advances, it often comes with inflated requirements in terms of resources to meaningfully scale. 
Wireless networks are currently evolving to cater to emerging cyber-physical and real-time interactive systems, such as swarm robotics, self-driving cars, and smart Internet of Things. A fundamental shift in thinking is necessary to satisfy the pressing requirements for timely multimodal communication, autonomous decision-making, and efficient distributed processing. Goal-oriented semantic communication is a new paradigm that aims at redefining importance, timing, and effectiveness in future networked intelligent systems \cite{popovski2020semantic,kountouris2021semantics,Qin22arxiv,tolga21SP}. Leveraging a minimalist design approach, it has the potential of significantly improving network resource usage, energy consumption, and computational efficiency. Various attempts in this direction have been made in the past \cite{Carnap,Juba,bao2011,SemanticGame} without though leading to an elegant and insightful theory with immediate practical applications.

In this context, \emph{semantics of information} is a recently emerged measure of the significance and the usefulness of messages with respect to the goal of data exchange. This composite performance metric appears to be instrumental in enabling effective communication of concise information that is both timely and valuable for achieving end users’ requirements. 
Age of information (AoI) performance metrics \cite{NowAoI,yates2021age}, which describe information freshness in networks, and value of information (VoI) \cite{VoI_USSR,VoI}, which quantifies the information utility or gain in decision making, can be viewed as simple, quantitative surrogate for information semantics. 

In this paper, we consider a communication system in which a transmitter receives status updates generated from a known discrete distribution with finite support and seeks to communicate them to a remote receiver. The updates generated by the information source may correspond to observations or measurements of a random phenomenon. 
The transmitter performs semantics-aware filtering and sends to the receiver only the most relevant randomly arriving source symbols in a timely fashion over an error-free channel. 
We consider a simple coding scheme focusing on less frequent events, i.e., the transmitter encodes only a fraction of the least frequent realizations, treating the remaining ones as not informative or irrelevant, thus providing more information about events that happen less often. Additionally, the semantics of information is captured through a timeliness metric for the received updates, which is a nonlinear function of age of information. Our objective is to design a coding scheme that optimizes the weighted sum of a semantics-aware utility function and a quadratic cost term on the average codeword length.     

This work falls within the realm of timely source coding problem \cite{TSC1,TSC2,TSC3}. These works study the design of lossless source codes and block codes that minimize the average age in status update systems under different queuing theoretic considerations. The most closely related to our work is \cite{bastopcu2020optimal}, which considers a selective encoding mechanism at the transmitter for timely updates. The optimal real codeword lengths that minimize the average age at the receiver are derived therein. Our paper extends previous results in several ways. We introduce semantics-aware metrics, which quantify update packet importance and timeliness of information at the receiver. The latter is a nonlinear function of age and we derive the average timeliness expression for three indicative cases. Furthermore, we add a quasiarithmetic penalty term related to the average codeword length \cite{Campbell,Larmore}. We derive the optimal real codeword lengths that maximize a semantics-aware utility function and minimize a quadratic average length cost, highlighting the performance gains of semantics-aware filtering and source coding.

\section{System Model}\label{Section2}
\begin{figure}[t!]
    \centering
    \pstool[scale=0.4]{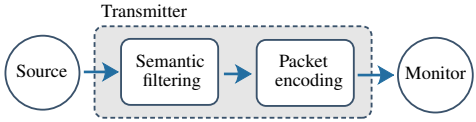}{
    \psfrag{Source}{\hspace{-0.12cm}\scriptsize Source}
    \psfrag{Semantic}{\hspace{-0.15cm}\scriptsize Semantic}
    \psfrag{Filtering}{\hspace{-0.106cm}\scriptsize filtering}
    \psfrag{Buffer}{\hspace{-0.15cm}\scriptsize Buffer}
    \psfrag{Selective}{\hspace{-0.06cm}\scriptsize Packet}
    \psfrag{Encoding}{\hspace{-0.106cm}\scriptsize encoding}
    \psfrag{FilteringP}{\hspace{-0.1cm}\scriptsize Filtering}
    \psfrag{olicy}{\hspace{-0.145cm}\scriptsize policy}
    \psfrag{ChanlStatusF}{\hspace{+0.1cm}\scriptsize Channel}
    \psfrag{eed}{\hspace{-0.25cm}\scriptsize condition}
    \psfrag{AccessN}{\hspace{-0.07cm}\scriptsize Access}
    \psfrag{etwork}{\hspace{-0.172cm}\scriptsize channel}
    \psfrag{Monitor}{\hspace{-0.16cm}\scriptsize Monitor}
    \psfrag{Transmitter}{\hspace{-0.0cm}\scriptsize Transmitter}
    }
    \vspace{-0.1cm}
    \caption{System model of semantics-aware transmission.}
    \label{Fig1}
\end{figure}

We consider a communication system in which an information source $X$ generates status updates in the form of packets and forwards them to a transmitter in order to send them to a remote monitor (c.f. Figure~\ref{Fig1}). The source generates discrete symbols from a finite set $\mathcal{X} = \{x_i ~ |~ i\!\in\!\mathcal{I}_n\}$, $\mathcal{I}_n = \{1,2,...,n\}$, each having a probability of realization $\tilde{p}_i = P_X(x_i)$ where $P_X(\cdot)$ is a known pmf (probability
mass function). Without loss of generality, we assume $\tilde{p}_i \geq \tilde{p}_j, \forall i \leq j$. Furthermore, we assume that the sequence of observations is independent and identically distributed and that packets are generated according to a Poisson distribution with rate $\lambda$.

Semantic filtering is performed, where only the $k$ least probable realizations are selected for transmission, while update packets from the remaining $n - k$ realizations are discarded. The set of selected update packets' indices (admitted packets) is denoted $\mathcal{I}_k \subseteq \mathcal{I}_n$. A first metric of semantic value associates importance with probability of occurrence of less frequent or atypical events. The less frequent an event (or the less probable a realization) is, the more important it is for the remote monitor. 
The transmitter then encodes an admitted packet from the $i$-th realization using a prefix-free code based on the truncated distribution with conditional probabilities $p_i = \tilde{p_i}/q_k$, $\forall i\in\mathcal{I}_k$ (and zero otherwise), where $q_k = \sum_{i\in \mathcal{I}_k} \tilde{p}_i$.

The transmitter node is bufferless, hence a newly admitted packet is blocked when the channel is busy. Assuming an error-free channel, if an admitted packet arrives at the transmitter when the channel is idle, it is correctly delivered to the receiver, then coined as a successive packet. 
After successfully delivering the previous packet to the receiver, the transmitter waits for the next admitted packet arrival.
We define $t_{i-1}$ the time instant that the $i$-th packet is received. Hence, the update interval between the $i$-th successive arrival and its next one is modeled as a random variable (r.v.) $Y_i = t_i - t_{i-1}$. This interval consists of the service time $S_i$ and the waiting time $W_i$. $W_i$ is the time between admitted status updates that are transmitted, thus the waiting time can be written as $W_i=\sum_{k=1}^{N} Z_k$, where $N$ is an r.v. of the number of admitted arrivals that are generated before finding the channel idle. $Z_k$ is the time between two admitted arrivals and is exponentially distributed with mean $\gamma = \frac{1}{\lambda q_k}$, since the admitted arrivals are generated according to a Poisson process with rate $\lambda q_k$.
The transmission time is proportional to the codeword length, thus, the service time (transmission time) of realization $x_i$ is $S_i = \ell_i$ time units, where $\ell_i$ is the length of the codeword assigned to $x_i$. The average transmission time is $\mathbb{E}[S]=\sum_{i=1}^{k} p_i \ell_i$.

\section{Problem Statement}
\subsection{Key Metrics of Interest}
We introduce a semantics of information (SoI) metric that measures the importance and usefulness of information at the receiver's side. SoI is generally a composite function $\mathcal{S}(t) = \nu(\psi(\mathcal{I})$, where $\psi:\mathbb{R}^m \to \mathbb{R}^z, m\geq z$ is a (nonlinear) function of $m \in \mathbb{Z}^+$ information attributes $\mathcal{I} \in \mathbb{R}^m$, and $\nu: \mathbb{R}^z \to \mathbb{R}$ is a context-dependent, cost-aware function \cite{kountouris2021semantics,pappas2021goal}. In this paper, we consider \textit{timeliness}, a contextual attribute defined as a non-increasing utility function $f:\mathbb{R}_0^+ \to \mathbb{R}$ of information freshness, i.e., $\mathcal{S}(t) = f(\Delta(t))$. $\Delta(t) = t - u(t)$ is the instantaneous AoI at the receiver, i.e., the difference of the current time instant and the timestamp $u(t)$ of the most recently received update. $\mathcal{S}(t)$ is a time varying stochastic process and the average SoI in stationary and ergodic systems for an observation interval $(0,T)$ is defined as $\bar{\mathcal{S}} = \displaystyle \underset{T\rightarrow\infty}{\lim} \dfrac{1}{T} \!\int_{0}^{T} f(\Delta(t)) {\rm d}t$.

\vspace{-2mm}
\subsection{Problem Formulation}
Our objective is to find the codeword lengths $\ell_i$ that optimize a weighted sum of the average SoI and the average length for a cost function $\phi(\ell):\mathbb{R}_0^+ \to \mathbb{R}_0^+$, i.e., $\sum_{i\in \mathcal{I}_k}p_i \phi(\ell_i)$. 
Maximizing the average SoI is equivalent to minimizing the average cost (penalty) of lateness
\begin{eqnarray}\label{Eq2}
L(\Delta) = \underset{T\rightarrow\infty}{\lim} \dfrac{1}{T} \!\int_{0}^{T} g(\Delta(t)) {\rm d}t
\end{eqnarray}
where $g:\mathbb{R}_0^+ \to \mathbb{R}$ is a non-decreasing function \cite{yates2021age}. Converting the maximization problem into a minimization one is mainly done for analytical convenience. 
The average codeword length term, also known as quasiarithmetic penalty, is related to Campbell's coding problem \cite{baer2006source}. 
The optimization problem is constrained by the integer constraint $\ell_i \in \mathbb{Z}^+$ and the Kraft-McMillan inequality \cite{cover1999elements} for the existence of a uniquely decodable code for a given set of codeword lengths.
Thus, we formulate the problem as  
\begin{equation}\label{Eq1}
\begin{aligned}
    \mathcal{P}_1\!:\,
    &\underset{{\{\ell_i\}}}{\text{min}}~~ L(\Delta) 
    + w \sum_{i\in \mathcal{I}_k} p_i \phi(\ell_i)
    \\
    &\text{s.t.} ~ \sum_{i\in \mathcal{I}_k} 2^{-\ell_i} \leq 1, \\
    &~~~~\, \ell_i \in \mathbb{Z}^+
\end{aligned}
\end{equation}
where $w>0$ denotes a weight parameter.
We employ a quadratic cost function for the codeword length under binary alphabetic $\phi(x) = \alpha x + \beta x^2$, ${\alpha}, {\beta} \geq 0$ \cite{Larmore}. Since $\phi$ is convex, longer (shorter) codewords are penalized more (less) harshly than in the linear case (e.g., Huffman coding) \cite{baer2006source}. 

First, we relax the integer constraint in $\mathcal{P}_1$ and allow non-negative real valued codeword lengths. Note that for any set of real-valued lengths $\ell_i$, we can obtain integer-valued lengths by using the rounded-off values $\lceil \ell_i \rceil$. 

\begin{figure}[t!]
    \centering
    \pstool[scale=0.53]{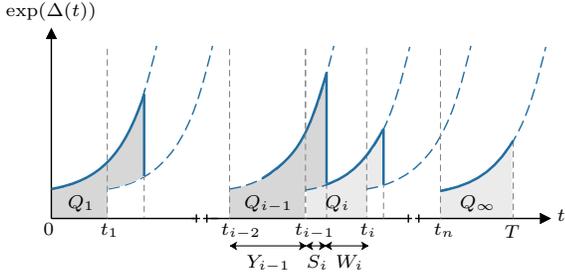}{
    \psfrag{U(Delta)}{\hspace{-0.32cm}\scriptsize $\operatorname{exp}(\Delta(t))$}
    \psfrag{t}{\hspace{0.05cm}\scriptsize $t$}
    \psfrag{0}{\hspace{0.0cm}\scriptsize $0$}
    \psfrag{t1}{\hspace{0.0cm}\scriptsize $t_{1}$}
    \psfrag{tn}{\hspace{0.0cm}\scriptsize $t_{n}$}
    \psfrag{ti-1}{\hspace{0.0cm}\scriptsize $t_{i-2}$}
    \psfrag{ti}{\hspace{-0.07cm}\scriptsize $t_{i-1}$}
    \psfrag{ti+1}{\hspace{0.05cm}\scriptsize $t_{i}$}
    \psfrag{Si}{\hspace{-0.05cm}\scriptsize $S_{i}$}
    \psfrag{Wi}{\hspace{-0.05cm}\scriptsize $W_{i}$}
    \psfrag{Yi+1}{\hspace{-0.1cm}\scriptsize $Y_{i-1}$}
    \psfrag{Q1}{\hspace{-0.06cm}\scriptsize $Q_{1}$}
    \psfrag{Qi}{\hspace{-0.2cm}\scriptsize $Q_{i-1}$}
    \psfrag{Qi+1}{\hspace{0.05cm}\scriptsize $Q_{i}$}
    \psfrag{Qn}{\hspace{-0.1cm}\scriptsize $Q_\infty$}
    \psfrag{Tau}{\hspace{0.02cm}\scriptsize $T$}
    \psfrag{umax}{\hspace{-0.06cm}\scriptsize $u_\text{max}$}
    \psfrag{Smax}{\hspace{-0.02cm}\scriptsize $S_\text{min}$}
    }
    \vspace{-0.1cm}
    \caption{Sample evolution of the exponential penalty function (\ref{Eq3}) over time for $\rho=1$.}
    \label{Fig2}
\end{figure}

Second, in order to explicitly calculate \eqref{Eq2}, we need to specify $g(\Delta(t))$. Three different instances of the penalty function are considered in this work. For exposition, in Figure \ref{Fig2} we show a sample path for the case $g(\Delta(t))={\rm exp}(\Delta(t))$. The calculation of \eqref{Eq2} is reduced to calculating the areas $Q_i$ in Figure \ref{Fig2} and then taking the average as follows
\begin{eqnarray}
    L(\Delta)=\underset{T\rightarrow\infty}{\lim} \dfrac{1}{T} \bigg\{\sum_{i=1}^{\mathcal{N}(T)} Q_i + {Q}_\infty \bigg\}\!
    = \eta \mathbb{E}[Q]
\end{eqnarray}
where $\eta = \underset{T\rightarrow\infty}{\lim}\frac{\mathcal{N}(T)\!-\!1}{T}$ is the steady-state time average arrival rate and $\mathcal{N}(T)$ is the number of admitted packet by time $T$. A more detailed and general analysis can be found in \cite{kosta2020cost}.

Merging $\eta$ with $w$ as both being positive constants, we have 
\begin{equation}\label{Eq6}
\begin{aligned}
    \mathcal{P}_2\!:\,
    &\underset{{\{\ell_i\}}}{\text{min}}~~ \underbrace{\mathbb{E}[Q] + w \sum_{i\in \mathcal{I}_k} p_i (\alpha \ell_i \!+\! \beta \ell_i^2)}_\text{$\triangleq \mathcal{J}_{{\rm SoI}}$}\\
    &\text{s.t.} ~ \sum_{i\in \mathcal{I}_k} 2^{-\ell_i} \leq 1, \\
    &~~~~\, \ell_i \in \mathbb{R}^+.
\end{aligned}
\end{equation}

\section{Semantics-aware Source Encoding Design}\label{Section3}
In this section, we determine the semantics-aware optimal real codeword lengths for three different instances of penalty function $g(\cdot)$, namely
\begin{align}\label{Eq3}
    g(\Delta(t))=
    \begin{cases}
    \operatorname{exp}(\rho\Delta(t))~~~\text{EDT case}\\
    \ln(\rho\Delta(t))~~~~~\text{LDT case}\\
   \rho(\Delta(t))^{\kappa}~~~~~~\text{PDT case}
    \end{cases}
\end{align}
where $\rho\geq0$ denotes a constant coefficient and $\kappa \in \mathbb{Z}^+$.
The above cases correspond to an \emph{exponentially} (E-), \emph{logarithmically} (L-), and \emph{polynomially} decreasing timeliness (PDT) scenario, respectively.

\subsection{Optimal Codeword Design}

\subsubsection{EDT Case} 
For this case, the area $Q_{i-1}$ for $i\geq3$ yields
\begin{eqnarray}\label{Eq7}\nonumber
Q_{i-1} \!\!\!\!&=&\!\!\!\! \int_{t_{i-2}}^{t_{i-1}+S_i} e^{\rho(t\!-\!t_{i-2})}{\rm d}t - \int_{t_{i-1}}^{t_{i-1}+S_i} e^{\rho (t-t_{i-1})}{\rm d}t\\\nonumber
\!\!\!\!&\stackrel{(a)}{\approx}&\!\!\!\! \dfrac{\rho}{2} Y_{i-1}^2 + \rho S_i Y_{i-1} + Y_{i-1}
\end{eqnarray}
where ($\alpha$) comes from using the second-order Taylor expansion for the exponential function.

Then, we calculate $\mathbb{E}[Q]$ as follows
\begin{align} \label{Eq8}
&\mathbb{E}[Q] \approx \dfrac{\rho}{2} \mathbb{E}[Y^2] + \rho \mathbb{E}[S] \mathbb{E}[Y] + \mathbb{E}[Y] \nonumber \\ 
&~~~\overset{(b)}{=} \dfrac{\rho}{2} \mathbb{E}[L^2] + \rho (\mathbb{E}[L])^2 + (1\!+\!2\rho\gamma) \mathbb{E}[L] + \rho\gamma^2 + \gamma.
\end{align}
To reach $(b)$, we have $\mathbb{E}[Y] = \mathbb{E}[L] + \gamma$, $\mathbb{E}[Y^2] = \mathbb{E}[L^2] + 2\gamma \mathbb{E}[L] + 2\gamma^2$, where $\gamma = (\lambda q_k)^{-1}$ \cite{bastopcu2020optimal}. Also, $\mathbb{E}[S] = \mathbb{E}[L]$ and $\mathbb{E}[S^2] = \mathbb{E}[L^2]$, with $\mathbb{E}[L] = \sum_{i\in \mathcal{I}_k} p_i \ell_i$, and $\mathbb{E}[L^2] = \sum_{i\in \mathcal{I}_k} p_i \ell_i^2$ being the first and second moments of the codeword lengths, respectively. 

Putting (\ref{Eq8}) into $\mathcal{P}_2$, we solve the following problem.
\begin{equation}\label{Eq9}
\begin{aligned}
    {\mathcal{P}}_3\!:\,
    &\underset{{\{\ell_i \in \mathbb{R}^+\}}}{\text{min}}\, \Big\{(\dfrac{\rho}{2} \!+\! w\beta) \mathbb{E}[L^2] + \rho (\mathbb{E}[L])^2  \\
    &~~~~~~~~~~ + (1\!+\!2\rho\gamma \!+\! w\alpha)\mathbb{E}[L] + \rho\gamma^2 + \gamma \Big\} \\
    &~~~\text{s.t.} ~ \sum_{i\in \mathcal{I}_k} 2^{-\ell_i} \leq 1.
\end{aligned}
\end{equation}

\begin{proposition}
The unique solution of problem $\mathcal{P}_3$ (EDT case) for $\ell_i$, $\forall i\in \mathcal{I}_k$, is given as
\begin{align}\label{Eq15}
    \ell_i = -\ln_2\!\left( \dfrac{\mathcal{C}_1 p_i}{\mu (\ln(2))^2} W_0\!\!\left(\! \dfrac{\mu (\ln(2))^2}{\mathcal{C}_1 p_i} 2^{\frac{\mathcal{C}_2}{\mathcal{C}_1}}\!\right) \!\right)\!
\end{align}
where $\mu\geq0$ is the Lagrange multiplier, $\mathcal{C}_1 = \rho + 2w\beta$,  
\begin{align}
    \mathcal{C}_2 = \frac{2\rho\mu \ln(2) + \mathcal{C}_1(1\!+\!2\rho\gamma \!+\! w\alpha)}{\mathcal{C}_1 + 2\rho},
\end{align}
and $W_0(.)$ is the principal branch of Lambert $W$ function.
\end{proposition}

\begin{IEEEproof}
We define the Lagrange function 
\begin{align}\label{Eq10}\nonumber
    \mathcal{L}(\ell_i;\mu) &= (\dfrac{\rho}{2} \!+\!w\beta) \sum_{i\in \mathcal{I}_k} p_i \ell_i^2 + \rho \bigg(\!\sum_{i\in \mathcal{I}_k} p_i \ell_i\!\bigg)^{\!\!2} \\ \nonumber
    &~~~+(1\!+\!2\rho\gamma \!+\! w\alpha)\bigg(\!\sum_{i\in \mathcal{I}_k} p_i \ell_i\!\bigg) + \rho\gamma^2 \\
    &~~~ + \gamma + \mu \bigg(\!\sum_{i\in \mathcal{I}_k} 2^{-\ell_i}\!-\!1\!\bigg)
\end{align}
where $\mu\geq0$ denotes the Lagrange multiplier. Now, we write the Karush-Kuhn-Tucker (KKT) condition as follows
\begin{align}\label{Eq11}\nonumber
    &\!\frac{\partial \mathcal{L}(\ell_i;\mu)}{\partial \ell_i} = (\rho \!+\! 2w\beta) p_i \ell_i +2\rho p_i \bigg(\!\sum_{i\in \mathcal{I}_k} p_i \ell_i\!\bigg) \\
    &~~~+ (1\!+\!2\rho\gamma \!+\! w\alpha)p_i - \mu \ln(2) 2^{-\ell_i}=0,~ \forall i\in\mathcal{I}_k.
\end{align}
The complementary slackness condition is 
\begin{align}\label{Eq12}
    \mu \bigg(\!\sum_{i\in \mathcal{I}_k} 2^{-\ell_i}\!-\!1\!\bigg)\!=0.
\end{align}
There exist two conditions, one of which meets (\ref{Eq12}): (i) $\mu=0$, hence $\sum_{i\in \mathcal{I}_k} 2^{-\ell_i} < 1$, or (ii) $\mu \neq 0$, hence $\sum_{i\in \mathcal{I}_k} 2^{-\ell_i}=1$. Condition (i) results in $\ell_i = \mathbb{E}[L] = - \big(\! \frac{1+2\rho\gamma + w\alpha}{3\rho + 2w\beta} \!\big) < 0$ from (\ref{Eq11}), which is not feasible. Thus, condition (ii) must satisfy (\ref{Eq12}). Thus, the moments of codeword lengths are obtained as
\begin{subequations}
\begin{align}\label{Eq13a}
    \mathbb{E}[L] &= \bigg(\!\frac{\mu \ln(2) -(1\!+\!2\rho\gamma\!+\! w\alpha)}{\mathcal{C}_1 + 2\rho}\!\bigg), \\ \label{Eq13b}
    \mathbb{E}[L^2] &= \bigg(\!\frac{\mu \ln(2) -(1\!+\!2\rho\gamma\!+\! w\alpha)}{\mathcal{C}_1 + 2\rho}\!\bigg)^{\!\!2}
\end{align}
\end{subequations}
where $\mathcal{C}_1 = \rho + 2w\beta$. Dividing (\ref{Eq11}) by $p_i$ and after some algebraic manipulations, we reach the following equation
\begin{align}\label{Eq14}
    \dfrac{\mu (\ln(2))^2}{\mathcal{C}_1 p_i}2^{- \ell_i} \operatorname{exp}\!\left(\!\dfrac{\mu (\ln(2))^2}{\mathcal{C}_1 p_i} 2^{-\ell_i}\!\right) = \dfrac{\mu (\ln(2))^2}{\mathcal{C}_1 p_i} 2^{\frac{\mathcal{C}_2}{\mathcal{C}_1}}
\end{align}
where $\mathcal{C}_2 = \frac{2\rho\mu \ln(2) + \mathcal{C}_1(1\!+\!2\rho\gamma \!+\! w\alpha)}{\mathcal{C}_1 + 2\rho}$.

The form of (\ref{Eq14}) is equal to $x \operatorname{exp}(x) = y$ for which the solution is $x = W_m(y)$, where $m=0$ or $m=\!-1$ if $y\geq0$ or $-e^{-1}\leq y<0$, respectively. 
\end{IEEEproof}
In order to find the optimal codeword lengths, we start from a value of $\mu$ that satisfies $\sum_{i\in \mathcal{I}_k} 2^{-\ell_i}=1$. Then, its value is updated by the use of (\ref{Eq13a}) and (\ref{Eq15}).

\subsubsection{LDT Case} 
In this case, the area $Q_{i-1}$ for $i\geq3$ yields
\begin{eqnarray*}
Q_{i-1} &=&\int_{t_{i-2}}^{t_{i-1}+S_i} \ln(\rho(t\!-\!t_{i-2})){\rm d}t\\
&-& \int_{t_{i-1}}^{t_{i-1}+S_i} \ln({\rho (t-t_{i-1})}){\rm d}t \\ &\approx& \rho Y_{i-1}^2 + 2\rho S_iY_{i-1} - 2Y_i,
\end{eqnarray*}
which results in
\begin{align}\label{Eq17}
    \mathbb{E}[Q] 
    = \rho \mathbb{E}[L^2] + 2\rho (\mathbb{E}[L])^2 + 2(2\rho\gamma\!-\!1) \mathbb{E}[L] + 2\rho\gamma^2 - 2\gamma.
\end{align}
Inserting (\ref{Eq17}) into $\mathcal{P}_2$, we obtain the following problem.
\begin{equation}\label{Eq18}
\begin{aligned}
    {\mathcal{P}}_4\!:\,
    &\underset{{\{\ell_i \in \mathbb{R}^+\}}}{\text{min}}\, \Big\{ (\rho \!+\! w\beta) \mathbb{E}[L^2] + 2\rho (\mathbb{E}[L])^2 \\
    &~~~~~~~~~~~~  + 2(2\rho\gamma\!-\!1\!+\!\dfrac{w\alpha}{2}) \mathbb{E}[L] + 2\rho\gamma^2 - 2\gamma  \Big\} \\
    &~~\text{s.t.} ~ \sum_{i\in \mathcal{I}_k} 2^{-\ell_i} \leq 1.
\end{aligned}
\end{equation}
Following the same procedure as (\ref{Eq10})--(\ref{Eq12}) and (\ref{Eq14}), the unique solution for $\ell_i$, for fixed $k$ is
\begin{align*}
    \ell_i = -\ln_2\!\left( \dfrac{\mathcal{C}_3 p_i}{\mu^{\prime} (\ln(2))^2} W_{0}\!\!\left(\!\dfrac{\mu^{\prime} (\ln(2))^2}{\mathcal{C}_3 p_i} 2^{\frac{\mathcal{C}_4}{\mathcal{C}_3}}\!\right) \!\right)\!
\end{align*}
where $\mu^{\prime}>0$, $\mathcal{C}_3 = 2\rho + 2w\beta$, and 
\begin{align*}
    \mathcal{C}_4 = \frac{4\rho\mu^\prime \ln(2) + 2\mathcal{C}_3(2\rho\gamma \!-\!1\!+\! \frac{w\alpha}{2})}{\mathcal{C}_3 + 4\rho}.
\end{align*}
Besides, the moments of the codeword lengths are given by
\begin{subequations}
\begin{align*}
    \mathbb{E}[L] &= \bigg(\!\frac{\mu^{\prime} \ln(2) -2(2\rho\gamma \!-\!1\!+\! \frac{w\alpha}{2})}{\mathcal{C}_3 + 4\rho}\!\bigg), \\ 
    \mathbb{E}[L^2] &= \bigg(\!\frac{\mu^{\prime} \ln(2) -2(2\rho\gamma \!-\!1\!+\! \frac{w\alpha}{2})}{\mathcal{C}_3 + 4\rho}\!\bigg)^{\!\!2}.
\end{align*}
\end{subequations}

\subsubsection{PDT Case} 
For this case (considering $\kappa = 1$), we obtain $Q_i = \frac{\rho}{2} Y_i^2 + \rho S_iY_{i-1}$, whose expected value is given by 
\begin{align}\label{Eq21}
    \mathbb{E}[Q] 
    = \dfrac{\rho}{2} \mathbb{E}[L^2] + \rho (\mathbb{E}[L])^2 + 2\rho\gamma \mathbb{E}[L] + \rho\gamma^2.
\end{align}

The resulting cost minimization problem (inserting (\ref{Eq21}) into $\mathcal{P}_2$) is then
\begin{equation}\label{Eq22}
\begin{aligned}
    {\mathcal{P}}_5\!:\,
    &\underset{{\{\ell_i \in \mathbb{R}^+\}}}{\text{min}}\, \Big\{ (\dfrac{\rho}{2} \!+\! w\beta) \mathbb{E}[L^2] + \rho (\mathbb{E}[L])^2 \\
    &~~~~~~~~~~~~  + (2\rho\gamma\!+\!{w\alpha}) \mathbb{E}[L] + \rho\gamma^2 \Big\} \\
    &~~\text{s.t.} ~ \sum_{i\in \mathcal{I}_k} 2^{-\ell_i} \leq 1.
\end{aligned}
\end{equation}
Similarly to other scenarios, the unique solution is  
\begin{align*}
    \ell_i = -\ln_2\!\left( \dfrac{\mathcal{C}_1 p_i}{\mu^{\prime\prime} (\ln(2))^2} W_{0}\!\!\left(\!\dfrac{\mu^{\prime\prime} (\ln(2))^2}{\mathcal{C}_1 p_i} 2^{\frac{\mathcal{C}_5}{\mathcal{C}_1}}\!\right) \!\right)\!
\end{align*}
where $\mu^{\prime\prime}>0$ and 
\begin{align*}
    \mathcal{C}_5 = \frac{2\rho\mu^{\prime\prime} \ln(2) + \mathcal{C}_1(2\rho\gamma \!+\! w\alpha)}{\mathcal{C}_1 + 2\rho}.
\end{align*}
The corresponding moments of codeword lengths are 
\begin{subequations}
\begin{align*}
    \mathbb{E}[L] &= \bigg(\!\frac{\mu^{\prime\prime} \ln(2) -(2\rho\gamma\!+\! {w\alpha})}{\mathcal{C}_1 + 2\rho}\!\bigg), \\ 
    \mathbb{E}[L^2] &= \bigg(\!\frac{\mu^{\prime\prime} \ln(2) -(2\rho\gamma\!+\! {w\alpha})}{\mathcal{C}_1 + 2\rho}\!\bigg)^{\!\!2}.
\end{align*}
\end{subequations}

\section{Numerical Results}\label{Section4}
In this section, we present numerical results in order to find SoI-optimal codeword lengths and to assess the performance gains of semantics-aware filtering and source coding. Unless otherwise stated, we use a Zipf($n,s$) distribution with pmf
\begin{eqnarray}
P_X(x)= \frac{1/x^s}{\sum_{j=1}^n 1/j^s},
\end{eqnarray}
with $n = \lvert \mathcal{X} \rvert = 100$ and the exponent $s = 0.4$. The parameter $s$ of the Zipf distribution allows us to vary from a uniform distribution ($s=0$) to a “peaky distribution”. We set $\rho=0.5$, ${\alpha} = {\beta} = 1$ and $T = 10\,[\text{sec}]$. For each scenario, the weight $w$ in the objective function is set in a way that the value range of average SoI and coding cost penalty terms becomes comparable. 
\begin{figure}[t!]
\centering
\subfloat[]{
\pstool[scale=0.55]{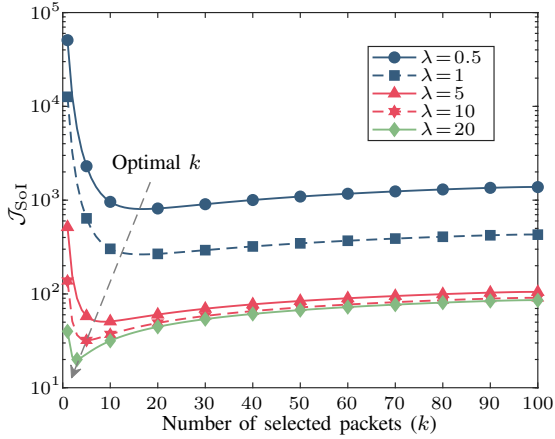}{
\psfrag{NormalizedSoI}{\hspace{0.35cm}\footnotesize $\mathcal{J}_{{\rm SoI}}$}
\psfrag{KofNsamples}{\hspace{-1.21cm}\footnotesize Number of selected packets ($k$)}
\psfrag{lambda=0.5}{\hspace{-0.0cm}\scriptsize $\lambda\!=\!0.5$}
\psfrag{lambda=1}{\hspace{-0.0cm}\scriptsize $\lambda\!=\!1$}
\psfrag{lambda=5}{\hspace{-0.0cm}\scriptsize $\lambda\!=\!5$}
\psfrag{lambda=10}{\hspace{-0.0cm}\scriptsize $\lambda\!=\!10$}
\psfrag{lambda=20}{\hspace{-0.0cm}\scriptsize $\lambda\!=\!20$}
\psfrag{Optimalk}{\hspace{-0.17cm}\footnotesize Optimal $k$}
\psfrag{0}{\hspace{-0.01cm}\scriptsize $0$}
\psfrag{10}{\hspace{-0.05cm}\scriptsize $10$}
\psfrag{20}{\hspace{-0.05cm}\scriptsize $20$}
\psfrag{30}{\hspace{-0.05cm}\scriptsize $30$}
\psfrag{40}{\hspace{-0.05cm}\scriptsize $40$}
\psfrag{50}{\hspace{-0.05cm}\scriptsize $50$}
\psfrag{60}{\hspace{-0.05cm}\scriptsize $60$}
\psfrag{70}{\hspace{-0.05cm}\scriptsize $70$}
\psfrag{80}{\hspace{-0.05cm}\scriptsize $80$}
\psfrag{90}{\hspace{-0.05cm}\scriptsize $90$}
\psfrag{100}{\hspace{-0.06cm}\scriptsize $100$}
\psfrag{0.5}{\hspace{-0.06cm}\scriptsize $0.5$}
\psfrag{0.6}{\hspace{-0.06cm}\scriptsize $0.6$}
\psfrag{0.7}{\hspace{-0.06cm}\scriptsize $0.7$}
\psfrag{0.8}{\hspace{-0.06cm}\scriptsize $0.8$}
\psfrag{0.9}{\hspace{-0.06cm}\scriptsize $0.9$}
\psfrag{L01}{\hspace{-0.04cm}\scriptsize $10^1$}
\psfrag{L02}{\hspace{-0.04cm}\scriptsize $10^2$}
\psfrag{L03}{\hspace{-0.04cm}\scriptsize $10^3$}
\psfrag{L04}{\hspace{-0.04cm}\scriptsize $10^4$}
\psfrag{L05}{\hspace{-0.04cm}\scriptsize $10^5$}
}}
\hfil
\vspace{-0.05cm}
\subfloat[]{
\pstool[scale=0.55]{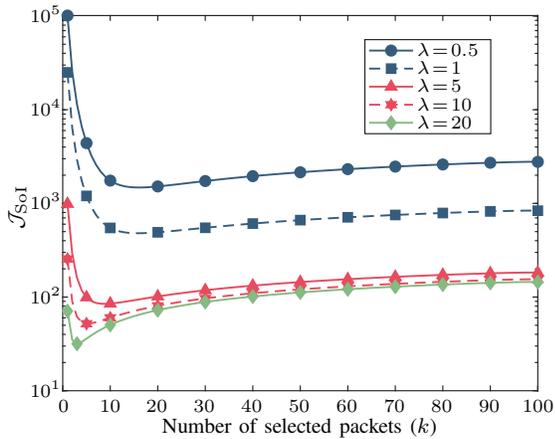}{
\psfrag{NormalizedSoI}{\hspace{0.35cm}\footnotesize $\mathcal{J}_{{\rm SoI}}$}
\psfrag{KofNsamples}{\hspace{-1.21cm}\footnotesize Number of selected packets ($k$)}
\psfrag{lambda=0.5}{\hspace{-0.0cm}\scriptsize $\lambda\!=\!0.5$}
\psfrag{lambda=1}{\hspace{-0.0cm}\scriptsize $\lambda\!=\!1$}
\psfrag{lambda=5}{\hspace{-0.0cm}\scriptsize $\lambda\!=\!5$}
\psfrag{lambda=10}{\hspace{-0.0cm}\scriptsize $\lambda\!=\!10$}
\psfrag{lambda=20}{\hspace{-0.0cm}\scriptsize $\lambda\!=\!20$}
\psfrag{0}{\hspace{-0.01cm}\scriptsize $0$}
\psfrag{10}{\hspace{-0.05cm}\scriptsize $10$}
\psfrag{20}{\hspace{-0.05cm}\scriptsize $20$}
\psfrag{30}{\hspace{-0.05cm}\scriptsize $30$}
\psfrag{40}{\hspace{-0.05cm}\scriptsize $40$}
\psfrag{50}{\hspace{-0.05cm}\scriptsize $50$}
\psfrag{60}{\hspace{-0.05cm}\scriptsize $60$}
\psfrag{70}{\hspace{-0.05cm}\scriptsize $70$}
\psfrag{80}{\hspace{-0.05cm}\scriptsize $80$}
\psfrag{90}{\hspace{-0.05cm}\scriptsize $90$}
\psfrag{100}{\hspace{-0.06cm}\scriptsize $100$}
\psfrag{0.5}{\hspace{-0.06cm}\scriptsize $0.5$}
\psfrag{0.6}{\hspace{-0.06cm}\scriptsize $0.6$}
\psfrag{0.7}{\hspace{-0.06cm}\scriptsize $0.7$}
\psfrag{0.8}{\hspace{-0.06cm}\scriptsize $0.8$}
\psfrag{0.9}{\hspace{-0.06cm}\scriptsize $0.9$}
\psfrag{L01}{\hspace{-0.04cm}\scriptsize $10^1$}
\psfrag{L02}{\hspace{-0.04cm}\scriptsize $10^2$}
\psfrag{L03}{\hspace{-0.04cm}\scriptsize $10^3$}
\psfrag{L04}{\hspace{-0.04cm}\scriptsize $10^4$}
\psfrag{L05}{\hspace{-0.04cm}\scriptsize $10^5$}
}}
\hfil
\vspace{-0.05cm}
\subfloat[]{
\pstool[scale=0.55]{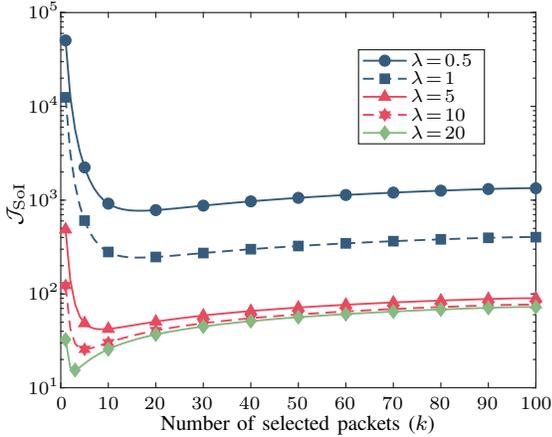}{
\psfrag{NormalizedSoI}{\hspace{0.35cm}\footnotesize $\mathcal{J}_{{\rm SoI}}$}
\psfrag{KofNsamples}{\hspace{-1.21cm}\footnotesize Number of selected packets ($k$)}
\psfrag{lambda=0.5}{\hspace{-0.0cm}\scriptsize $\lambda\!=\!0.5$}
\psfrag{lambda=1}{\hspace{-0.0cm}\scriptsize $\lambda\!=\!1$}
\psfrag{lambda=5}{\hspace{-0.0cm}\scriptsize $\lambda\!=\!5$}
\psfrag{lambda=10}{\hspace{-0.0cm}\scriptsize $\lambda\!=\!10$}
\psfrag{lambda=20}{\hspace{-0.0cm}\scriptsize $\lambda\!=\!20$}
\psfrag{0}{\hspace{-0.01cm}\scriptsize $0$}
\psfrag{10}{\hspace{-0.05cm}\scriptsize $10$}
\psfrag{20}{\hspace{-0.05cm}\scriptsize $20$}
\psfrag{30}{\hspace{-0.05cm}\scriptsize $30$}
\psfrag{40}{\hspace{-0.05cm}\scriptsize $40$}
\psfrag{50}{\hspace{-0.05cm}\scriptsize $50$}
\psfrag{60}{\hspace{-0.05cm}\scriptsize $60$}
\psfrag{70}{\hspace{-0.05cm}\scriptsize $70$}
\psfrag{80}{\hspace{-0.05cm}\scriptsize $80$}
\psfrag{90}{\hspace{-0.05cm}\scriptsize $90$}
\psfrag{100}{\hspace{-0.06cm}\scriptsize $100$}
\psfrag{0.5}{\hspace{-0.06cm}\scriptsize $0.5$}
\psfrag{0.6}{\hspace{-0.06cm}\scriptsize $0.6$}
\psfrag{0.7}{\hspace{-0.06cm}\scriptsize $0.7$}
\psfrag{0.8}{\hspace{-0.06cm}\scriptsize $0.8$}
\psfrag{0.9}{\hspace{-0.06cm}\scriptsize $0.9$}
\psfrag{L01}{\hspace{-0.04cm}\scriptsize $10^1$}
\psfrag{L02}{\hspace{-0.04cm}\scriptsize $10^2$}
\psfrag{L03}{\hspace{-0.04cm}\scriptsize $10^3$}
\psfrag{L04}{\hspace{-0.04cm}\scriptsize $10^4$}
\psfrag{L05}{\hspace{-0.04cm}\scriptsize $10^5$}
}}
\caption{The objective function $\mathcal{J}_{{\rm SoI}}$ versus the number of selected packets $k$ for the (a) EDT, (b) LDT, and (c) PDT scenarios with Zipf(100,0.4) distribution.}
\label{Fig3}
\end{figure}

Figures~\ref{Fig3}\,(a), \ref{Fig3}\,(b), and \ref{Fig3}\,(c) show the value of the objective function $\mathcal{J}_{{\rm SoI}}$ (i.e., cost of lateness and coding penalty term) versus the number of $k$ realizations for the EDT, LDT, and PDT cases, respectively. Evidently, increasing the arrival rate reduces $\mathcal{J}_{{\rm SoI}}$ as well as the optimal $k$. 
For infrequent update arrivals, the transmitter does not filter out most updates ($k \neq 1$), whereas no filtering ($k \to n$) results in performance degradation due to longer transmission times for infrequent realizations. Among the three sample cases, the PDT (LDT) scenario offers the lowest (highest) value of $\mathcal{J}_{{\rm SoI}}$. Comparing with the linear age scenario $g(\Delta(t)) \!=\! \rho\Delta(t)$, the optimal $k$ is $19.3$, $13.2$, $9.8$, $7$, and $5.3$ for the arrival rates of $0.5$, $1$, $5$, $10$, and $20$, respectively. Therefore, an exponential penalty (nonlinear age) results in lower values for optimal $k$ compared to the linear one (cf. Figure~\ref{Fig3}\,(a)).

\begin{figure}[t!]
\centering
\pstool[scale=0.55]{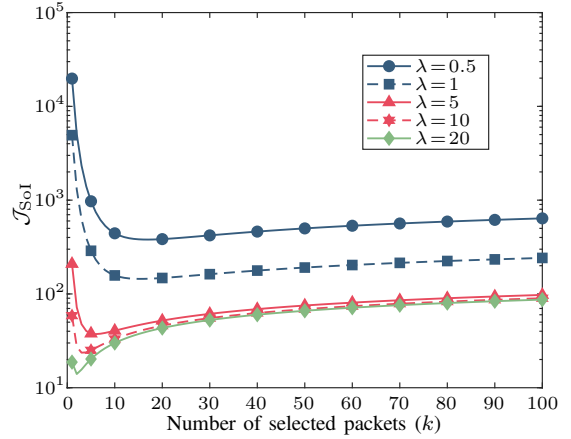}{
\psfrag{NormalizedSoI}{\hspace{0.35cm}\footnotesize $\mathcal{J}_{{\rm SoI}}$}
\psfrag{KofNsamples}{\hspace{-1.21cm}\footnotesize Number of selected packets ($k$)}
\psfrag{lambda=0.5}{\hspace{-0.0cm}\scriptsize $\lambda\!=\!0.5$}
\psfrag{lambda=1}{\hspace{-0.0cm}\scriptsize $\lambda\!=\!1$}
\psfrag{lambda=5}{\hspace{-0.0cm}\scriptsize $\lambda\!=\!5$}
\psfrag{lambda=10}{\hspace{-0.0cm}\scriptsize $\lambda\!=\!10$}
\psfrag{lambda=20}{\hspace{-0.0cm}\scriptsize $\lambda\!=\!20$}
\psfrag{0}{\hspace{-0.01cm}\scriptsize $0$}
\psfrag{10}{\hspace{-0.05cm}\scriptsize $10$}
\psfrag{20}{\hspace{-0.05cm}\scriptsize $20$}
\psfrag{30}{\hspace{-0.05cm}\scriptsize $30$}
\psfrag{40}{\hspace{-0.05cm}\scriptsize $40$}
\psfrag{50}{\hspace{-0.05cm}\scriptsize $50$}
\psfrag{60}{\hspace{-0.05cm}\scriptsize $60$}
\psfrag{70}{\hspace{-0.05cm}\scriptsize $70$}
\psfrag{80}{\hspace{-0.05cm}\scriptsize $80$}
\psfrag{90}{\hspace{-0.05cm}\scriptsize $90$}
\psfrag{100}{\hspace{-0.06cm}\scriptsize $100$}
\psfrag{0.5}{\hspace{-0.06cm}\scriptsize $0.5$}
\psfrag{0.6}{\hspace{-0.06cm}\scriptsize $0.6$}
\psfrag{0.7}{\hspace{-0.06cm}\scriptsize $0.7$}
\psfrag{0.8}{\hspace{-0.06cm}\scriptsize $0.8$}
\psfrag{0.9}{\hspace{-0.06cm}\scriptsize $0.9$}
\psfrag{L01}{\hspace{-0.04cm}\scriptsize $10^1$}
\psfrag{L02}{\hspace{-0.04cm}\scriptsize $10^2$}
\psfrag{L03}{\hspace{-0.04cm}\scriptsize $10^3$}
\psfrag{L04}{\hspace{-0.04cm}\scriptsize $10^4$}
\psfrag{L05}{\hspace{-0.04cm}\scriptsize $10^5$}
}
\caption{The objective function $\mathcal{J}_{{\rm SoI}}$ versus $k$ for uniform probability distributions under the EDT scenario and $n=100$.}
\label{Fig3new}
\end{figure}

To investigate the effect of the pmf and the source characteristics on the performance, in Figure~\ref{Fig3new} we plot the objective function $\mathcal{J}_{{\rm SoI}}$ versus $k$ for the EDT case under uniform distribution. Despite the similarity in the shape, the optimal $k$ is slightly smaller than the Zipf pmf. The reason is that the critical point of the objective function versus $k$, hence $q_k$, is proportional to the input rate. Thus, for each input rate, there is an optimal $q_k$ yielding an optimal value for $k$. For instance, for $\lambda=10$, the Zipf and the uniform distribution results in optimal $k=5$ and $k=3$, respectively. 

\begin{figure}[t!]
\centering
\subfloat[]{
\pstool[scale=0.55]{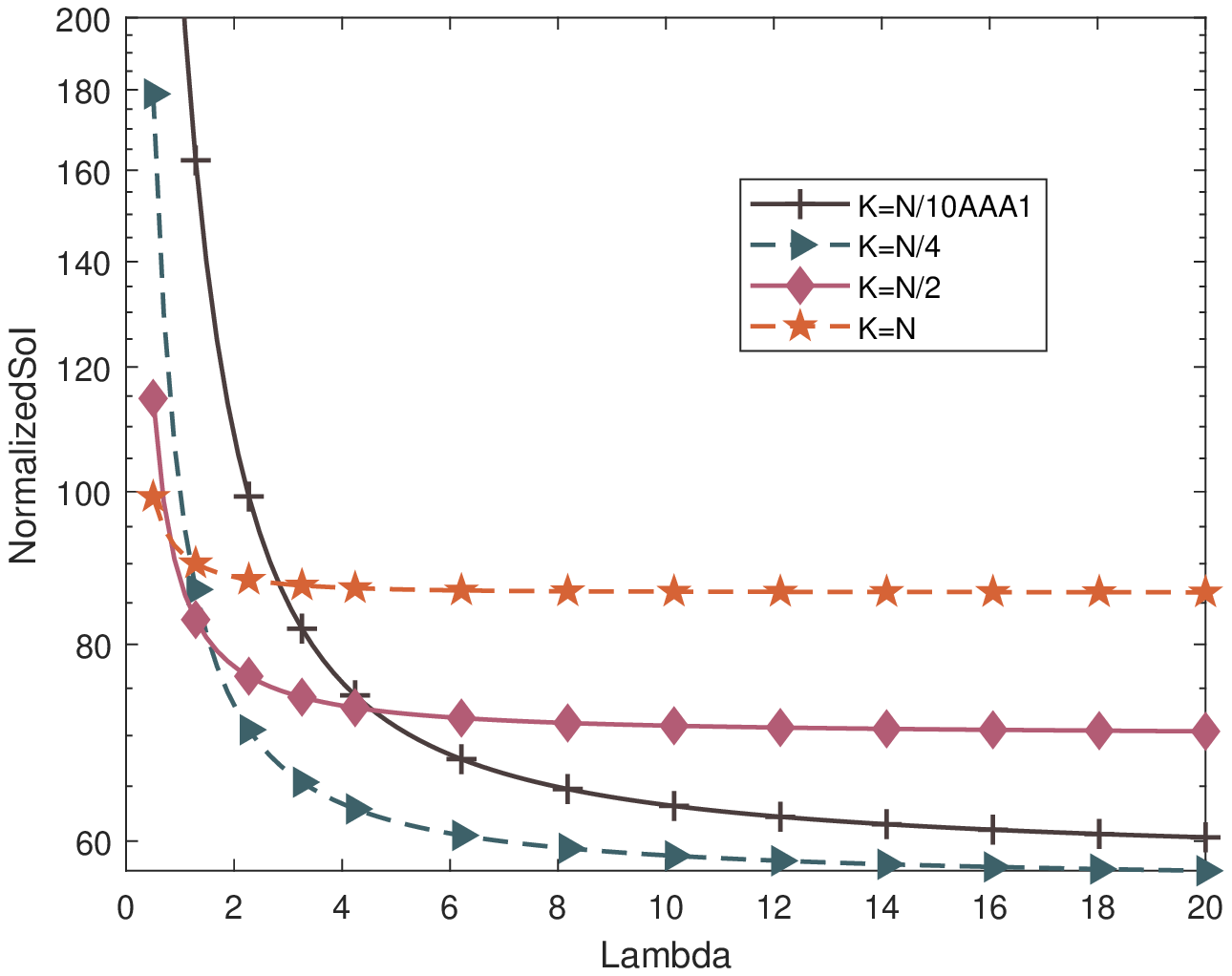}{
\psfrag{NormalizedSoI}{\hspace{0.35cm}\footnotesize $\mathcal{J}_{{\rm SoI}}$}
\psfrag{Lambda}{\hspace{-0.1cm}\footnotesize Rate ($\lambda$)}
\psfrag{K=N/10AAA1}{\hspace{-0.0cm}\scriptsize $k\!=\!n/10$}
\psfrag{K=N/4}{\hspace{-0.0cm}\scriptsize $k\!=\!n/4$}
\psfrag{K=N/2}{\hspace{-0.0cm}\scriptsize $k\!=\!n/2$}
\psfrag{K=N}{\hspace{-0.0cm}\scriptsize $k\!=\!n$}
\psfrag{0}{\hspace{-0.01cm}\scriptsize $0$}
\psfrag{0.5}{\hspace{-0.06cm}\scriptsize $0.5$}
\psfrag{0.6}{\hspace{-0.06cm}\scriptsize $0.6$}
\psfrag{0.7}{\hspace{-0.06cm}\scriptsize $0.7$}
\psfrag{0.8}{\hspace{-0.06cm}\scriptsize $0.8$}
\psfrag{0.9}{\hspace{-0.06cm}\scriptsize $0.9$}
\psfrag{1}{\hspace{-0.02cm}\scriptsize $1$}
\psfrag{2}{\hspace{-0.01cm}\scriptsize $2$}
\psfrag{4}{\hspace{-0.01cm}\scriptsize $4$}
\psfrag{6}{\hspace{-0.01cm}\scriptsize $6$}
\psfrag{8}{\hspace{-0.01cm}\scriptsize $8$}
\psfrag{10}{\hspace{-0.01cm}\scriptsize $10$}
\psfrag{12}{\hspace{-0.01cm}\scriptsize $12$}
\psfrag{14}{\hspace{-0.01cm}\scriptsize $14$}
\psfrag{16}{\hspace{-0.01cm}\scriptsize $16$}
\psfrag{18}{\hspace{-0.01cm}\scriptsize $18$}
\psfrag{20}{\hspace{-0.01cm}\scriptsize $20$}
\psfrag{60}{\hspace{-0.06cm}\scriptsize $60$}
\psfrag{80}{\hspace{-0.06cm}\scriptsize $80$}
\psfrag{100}{\hspace{-0.06cm}\scriptsize $100$}
\psfrag{120}{\hspace{-0.06cm}\scriptsize $120$}
\psfrag{140}{\hspace{-0.06cm}\scriptsize $140$}
\psfrag{160}{\hspace{-0.06cm}\scriptsize $160$}
\psfrag{180}{\hspace{-0.06cm}\scriptsize $180$}
\psfrag{200}{\hspace{-0.06cm}\scriptsize $200$}
}}
\hfil
\vspace{-0.05cm}
\subfloat[]{
\pstool[scale=0.55]{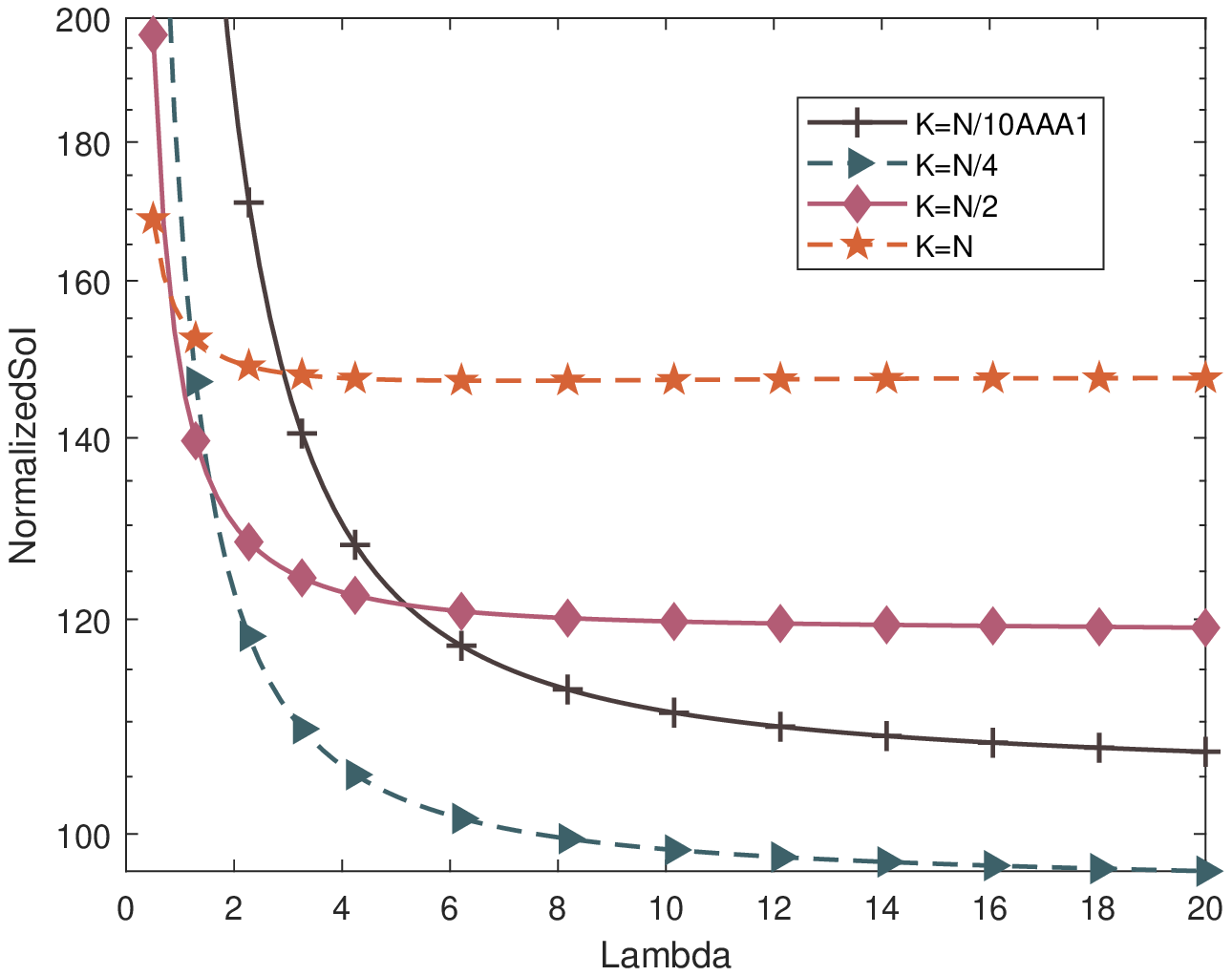}{
\psfrag{NormalizedSoI}{\hspace{0.35cm}\footnotesize $\mathcal{J}_{{\rm SoI}}$}
\psfrag{Lambda}{\hspace{-0.1cm}\footnotesize Rate ($\lambda$)}
\psfrag{K=N/10AAA1}{\hspace{-0.0cm}\scriptsize $k\!=\!n/10$}
\psfrag{K=N/4}{\hspace{-0.0cm}\scriptsize $k\!=\!n/4$}
\psfrag{K=N/2}{\hspace{-0.0cm}\scriptsize $k\!=\!n/2$}
\psfrag{K=N}{\hspace{-0.0cm}\scriptsize $k\!=\!n$}
\psfrag{0}{\hspace{-0.01cm}\scriptsize $0$}
\psfrag{0.5}{\hspace{-0.06cm}\scriptsize $0.5$}
\psfrag{0.6}{\hspace{-0.06cm}\scriptsize $0.6$}
\psfrag{0.7}{\hspace{-0.06cm}\scriptsize $0.7$}
\psfrag{0.8}{\hspace{-0.06cm}\scriptsize $0.8$}
\psfrag{0.9}{\hspace{-0.06cm}\scriptsize $0.9$}
\psfrag{1}{\hspace{-0.02cm}\scriptsize $1$}
\psfrag{2}{\hspace{-0.01cm}\scriptsize $2$}
\psfrag{4}{\hspace{-0.01cm}\scriptsize $4$}
\psfrag{6}{\hspace{-0.01cm}\scriptsize $6$}
\psfrag{8}{\hspace{-0.01cm}\scriptsize $8$}
\psfrag{10}{\hspace{-0.01cm}\scriptsize $10$}
\psfrag{12}{\hspace{-0.01cm}\scriptsize $12$}
\psfrag{14}{\hspace{-0.01cm}\scriptsize $14$}
\psfrag{16}{\hspace{-0.01cm}\scriptsize $16$}
\psfrag{18}{\hspace{-0.01cm}\scriptsize $18$}
\psfrag{20}{\hspace{-0.01cm}\scriptsize $20$}
\psfrag{100}{\hspace{-0.06cm}\scriptsize $100$}
\psfrag{120}{\hspace{-0.06cm}\scriptsize $120$}
\psfrag{140}{\hspace{-0.06cm}\scriptsize $140$}
\psfrag{160}{\hspace{-0.06cm}\scriptsize $160$}
\psfrag{180}{\hspace{-0.06cm}\scriptsize $180$}
\psfrag{200}{\hspace{-0.06cm}\scriptsize $200$}
}}
\hfil
\vspace{-0.05cm}
\subfloat[]{
\pstool[scale=0.55]{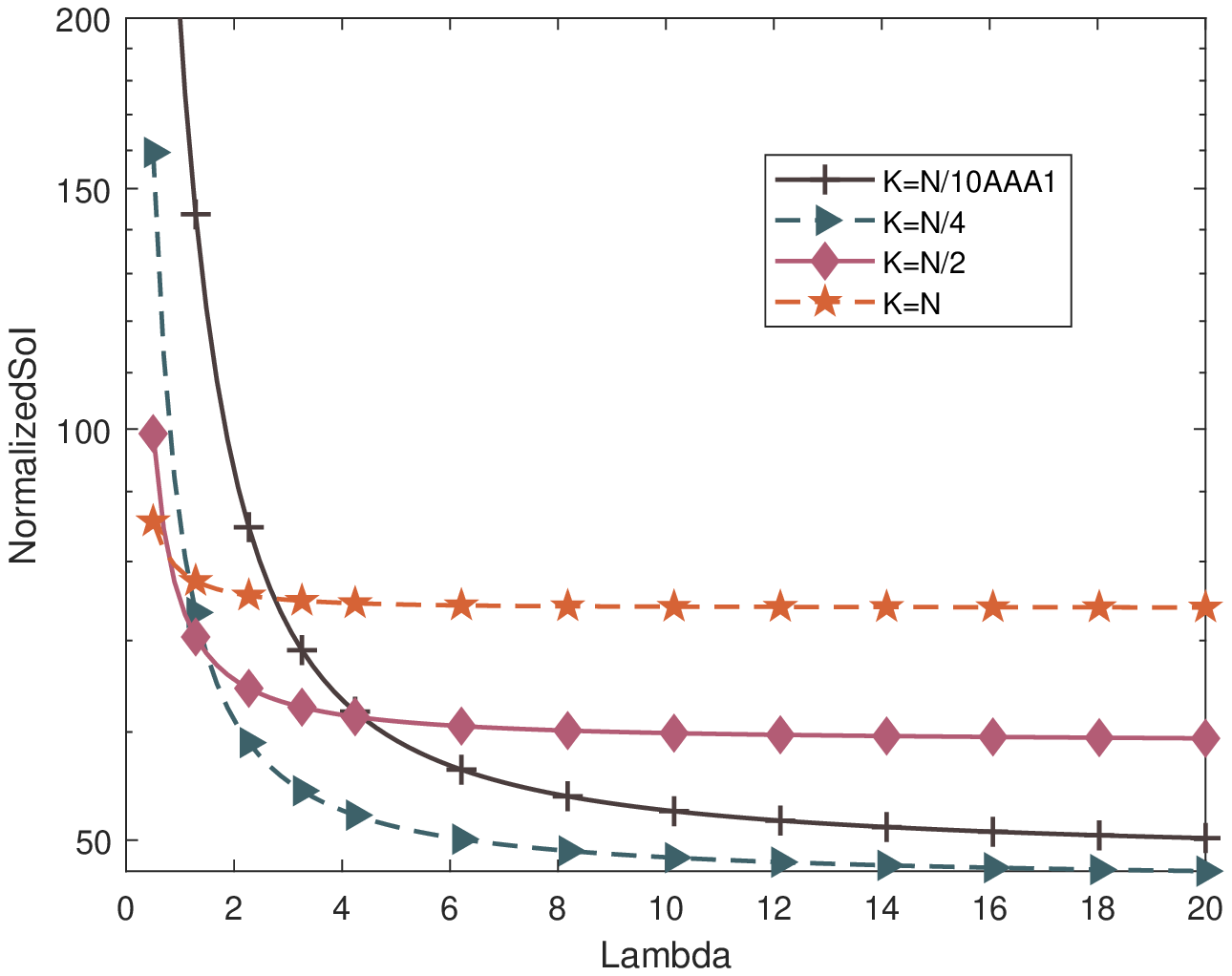}{
\psfrag{NormalizedSoI}{\hspace{0.35cm}\footnotesize $\mathcal{J}_{{\rm SoI}}$}
\psfrag{Lambda}{\hspace{-0.1cm}\footnotesize Rate ($\lambda$)}
\psfrag{K=N/10AAA1}{\hspace{-0.0cm}\scriptsize $k\!=\!n/10$}
\psfrag{K=N/4}{\hspace{-0.0cm}\scriptsize $k\!=\!n/4$}
\psfrag{K=N/2}{\hspace{-0.0cm}\scriptsize $k\!=\!n/2$}
\psfrag{K=N}{\hspace{-0.0cm}\scriptsize $k\!=\!n$}
\psfrag{0}{\hspace{-0.01cm}\scriptsize $0$}
\psfrag{0.5}{\hspace{-0.06cm}\scriptsize $0.5$}
\psfrag{0.6}{\hspace{-0.06cm}\scriptsize $0.6$}
\psfrag{0.7}{\hspace{-0.06cm}\scriptsize $0.7$}
\psfrag{0.8}{\hspace{-0.06cm}\scriptsize $0.8$}
\psfrag{0.9}{\hspace{-0.06cm}\scriptsize $0.9$}
\psfrag{1}{\hspace{-0.02cm}\scriptsize $1$}
\psfrag{2}{\hspace{-0.01cm}\scriptsize $2$}
\psfrag{4}{\hspace{-0.01cm}\scriptsize $4$}
\psfrag{6}{\hspace{-0.01cm}\scriptsize $6$}
\psfrag{8}{\hspace{-0.01cm}\scriptsize $8$}
\psfrag{10}{\hspace{-0.01cm}\scriptsize $10$}
\psfrag{12}{\hspace{-0.01cm}\scriptsize $12$}
\psfrag{14}{\hspace{-0.01cm}\scriptsize $14$}
\psfrag{16}{\hspace{-0.01cm}\scriptsize $16$}
\psfrag{18}{\hspace{-0.01cm}\scriptsize $18$}
\psfrag{20}{\hspace{-0.01cm}\scriptsize $20$}
\psfrag{50}{\hspace{-0.06cm}\scriptsize $50$}
\psfrag{100}{\hspace{-0.06cm}\scriptsize $100$}
\psfrag{150}{\hspace{-0.06cm}\scriptsize $150$}
\psfrag{200}{\hspace{-0.06cm}\scriptsize $200$}
}}
\caption{The objective function $\mathcal{J}_{{\rm SoI}}$ versus rate $\lambda$ for the (a) EDT, (b) LDT, and (c) PDT scenarios and Zipf(100,0.4) pmf.}
\label{Fig4}
\end{figure}

Figure~\ref{Fig4} depicts the objective function versus the rate parameter $\lambda$ for different values of $k$. Increasing the input rate decreases $\mathcal{J}_{\rm SoI}$; however, this decrease diminishes and saturates at higher rate values. Furthermore, by increasing the number of selected packets, lower input rates are required to reduce the penalty terms. For instance, in the EDT scenario, the lowest attained $\mathcal{J}_{\rm SoI}$ value is $60$, $57$, $71$, and $87$ for $k=\frac{n}{10}$, $\frac{n}{4}$, $\frac{n}{2}$, and $n$, respectively. For large $k$, the objective function gets high values for any input rates. Based on the analytical expressions derived throughout the paper, in the EDT case, we find the global optimal values of $\lambda^* = 19.34$, $16.71$, $10.12$, and $5.83$ for $k=\frac{n}{10}$, $\frac{n}{4}$, $\frac{n}{2}$, and $n$, respectively.

\begin{figure}[t!]
\centering
\pstool[scale=0.55]{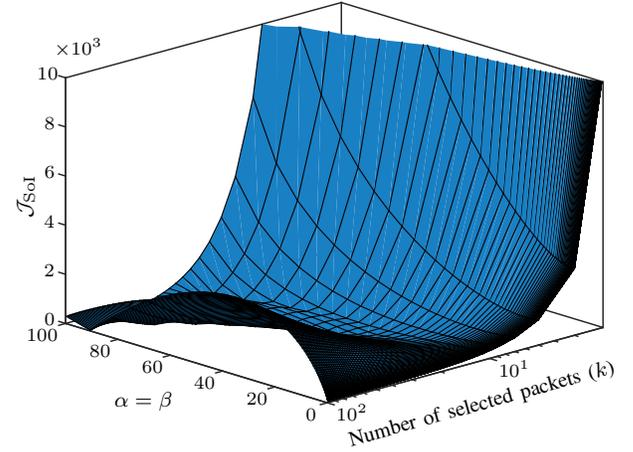}{
\psfrag{NormalizedSoI}{\hspace{0.3cm}\footnotesize $\mathcal{J}_{{\rm SoI}}$}
\psfrag{KofNsamples}{\hspace{-1.35cm}\footnotesize \rotatebox{15.55}{Number of selected packets ($k$)}}
\psfrag{AlphaBeta}{\hspace{0cm}\footnotesize ${\alpha}={\beta}$}
\psfrag{K=N/10AAA1}{\hspace{-0.0cm}\scriptsize $k\!=\!n/10$}
\psfrag{K=N/4}{\hspace{-0.0cm}\scriptsize $k\!=\!n/4$}
\psfrag{K=N/2}{\hspace{-0.0cm}\scriptsize $k\!=\!n/2$}
\psfrag{K=N}{\hspace{-0.0cm}\scriptsize $k\!=\!n$}
\psfrag{0}{\hspace{-0.02cm}\scriptsize $0$}
\psfrag{2}{\hspace{-0.02cm}\scriptsize $2$}
\psfrag{4}{\hspace{-0.02cm}\scriptsize $4$}
\psfrag{6}{\hspace{-0.02cm}\scriptsize $6$}
\psfrag{8}{\hspace{-0.02cm}\scriptsize $8$}
\psfrag{10}{\hspace{-0.02cm}\scriptsize $10$}
\psfrag{20}{\hspace{-0.01cm}\scriptsize $20$}
\psfrag{40}{\hspace{-0.01cm}\scriptsize $40$}
\psfrag{60}{\hspace{-0.01cm}\scriptsize $60$}
\psfrag{80}{\hspace{-0.01cm}\scriptsize $80$}
\psfrag{100}{\hspace{-0.06cm}\scriptsize $100$}
\psfrag{L01}{\hspace{-0.06cm}\scriptsize $10^1$}
\psfrag{L02}{\hspace{-0.06cm}\scriptsize $10^2$}
\psfrag{10x3}{\hspace{0.15cm}\scriptsize $\times 10^3$}
}
\vspace{-0.5cm}
\caption{The interplay among $\mathcal{J}_{{\rm SoI}}$, selected packets $k$ and codeword length cost parameters $\alpha, \beta$ in the EDT scenario with $n=100$ and $\lambda=1$.}
\label{Fig5}
\end{figure}

Figure~\ref{Fig5} plots the objective function $\mathcal{J}_{{\rm SoI}}$ versus the number of selected packets $k$ and the values of the cost parameters (i.e., ${\alpha}$, ${\beta}$) under the EDT scenario. Herein, the cost parameters are assumed to have equal values and $\lambda=1$. As expected, the optimal values of the coding cost parameters depend on the number of selected packets. The objective function continuously increases as the cost parameters increase for small $k$. However, for large $k$, an increase of the cost parameters causes the objective function to first increase and then decrease. The interplay between the two terms of the objective function (timeliness penalty and coding cost) and the number of selected packets $k$ is summarized in Table~\ref{Tab1}, which shows the optimal values of $k$, $\alpha$, and $\beta$ for different input rates under the EDT scenario. We observe that increasing the input rate, hence decreasing the optimal $k$, the optimal values of cost parameters increase. When the input rate is high, one has to assign larger penalties for the codeword length to ensure transmitting the most important data and allocating them larger codewords.

\noindent
\renewcommand{\arraystretch}{1}
\begin{table}[t!]
\begin{center}
\caption{Optimal parameters under the EDT scenario.}\label{Tab1}
\begin{tabular}{ | c | c | c || c | c | c |}
\hline
\small \!\!$\lambda$\!\! & \small \!$k$\! & \small \!$\alpha=\beta$\! & \small \!\!$\lambda$\!\! & \small \!$k$\! & \small \!$\alpha=\beta$\!\\
\hline
\hline
\small $0.5$ & \footnotesize $20$ & \footnotesize $1.26$ & \small $10$ & \footnotesize $5$ & \footnotesize $2.5$\\
\hline
\small $1$ & \footnotesize $18$ & \footnotesize $1.58$  & \small $20$ & \footnotesize $2$ & \footnotesize $12.59$\\
\hline
\small $5$ & \footnotesize $10$ & \footnotesize $1.99$\\
\cline{1-3}
\end{tabular}
\medskip
\end{center}
\end{table}

\section{Conclusion}\label{Section5}
We studied the problem of timely source coding in status update systems, where the transmitter selects the packets generated by an information source based on their importance prior sending them to a remote receiver. Introducing a semantics-aware metric that quantifies information timeliness, we determined the real codeword lengths that optimize a weighted sum of timeliness and quadratic coding cost penalty. The main takeaway is that semantic filtering and source coding can significantly reduce the number of packets that has to be communicated while providing timely updates.

\bibliographystyle{IEEEtran}
\bibliography{References.bib}

\begin{thebibliography}{10}
\providecommand{\url}[1]{#1}
\csname url@samestyle\endcsname
\providecommand{\newblock}{\relax}
\providecommand{\bibinfo}[2]{#2}
\providecommand{\BIBentrySTDinterwordspacing}{\spaceskip=0pt\relax}
\providecommand{\BIBentryALTinterwordstretchfactor}{4}
\providecommand{\BIBentryALTinterwordspacing}{\spaceskip=\fontdimen2\font plus
\BIBentryALTinterwordstretchfactor\fontdimen3\font minus
  \fontdimen4\font\relax}
\providecommand{\BIBforeignlanguage}[2]{{%
\expandafter\ifx\csname l@#1\endcsname\relax
\typeout{** WARNING: IEEEtran.bst: No hyphenation pattern has been}%
\typeout{** loaded for the language `#1'. Using the pattern for}%
\typeout{** the default language instead.}%
\else
\language=\csname l@#1\endcsname
\fi
#2}}
\providecommand{\BIBdecl}{\relax}
\BIBdecl

\bibitem{popovski2020semantic}
P.~Popovski, O.~Simeone, F.~Boccardi, D.~G{\"u}nd{\"u}z, and O.~Sahin,
  ``Semantic-effectiveness filtering and control for post-{5G} wireless
  connectivity,'' \emph{Journal of the Indian Institute of Science}, vol. 100,
  no.~2, pp. 435--443, 2020.

\bibitem{kountouris2021semantics}
M.~Kountouris and N.~Pappas, ``Semantics-empowered communication for networked
  intelligent systems,'' \emph{IEEE Communications Magazine}, vol.~59, no.~6,
  pp. 96--102, 2021.

\bibitem{Qin22arxiv}
Z.~Qin, X.~Tao, J.~Lu, and G.~Y. Li, ``Semantic communications: Principles and
  challenges,'' \emph{arXiv preprint arXiv: 2201.01389v2}, 2022.

\bibitem{tolga21SP}
M.~Kalfa, M.~Gok, A.~Atalik, B.~Tegin, T.~M. Duman, and O.~Arikan, ``Towards
  goal-oriented semantic signal processing: Applications and future
  challenges,'' \emph{Digit. Signal Process.}, vol. 119, 2021.

\bibitem{Carnap}
Y.~Bar{-}Hillel and R.~Carnap, ``Semantic information,'' \emph{British Journal
  for the Philosophy of Science}, vol.~4, no.~14, 1953.

\bibitem{Juba}
B.~Juba and M.~Sudan, ``Universal semantic communication {I},'' in \emph{Proc.
  of the 40th Annual ACM Symposium on Theory of Computing}, ser. STOC '08, New
  York, NY, USA, 2008, p. 123–132.

\bibitem{bao2011}
J.~{Bao}, P.~{Basu}, M.~{Dean}, C.~{Partridge}, A.~{Swami}, W.~{Leland}, and
  J.~A. {Hendler}, ``Towards a theory of semantic communication,'' in
  \emph{IEEE Network Science Workshop}, 2011, pp. 110--117.

\bibitem{SemanticGame}
B.~{Güler}, A.~{Yener}, and A.~{Swami}, ``The semantic communication game,''
  \emph{IEEE Trans. on Cog. Comm. and Net.}, vol.~4, no.~4, 2018.

\bibitem{NowAoI}
A.~Kosta, N.~Pappas, and V.~Angelakis, ``Age of information: A new concept,
  metric, and tool,'' \emph{Foundations and Trends in Networking}, vol.~12,
  no.~3, 2017.

\bibitem{yates2021age}
R.~D. Yates, Y.~Sun, D.~R. Brown, S.~K. Kaul, E.~Modiano, and S.~Ulukus, ``Age
  of information: An introduction and survey,'' \emph{IEEE Journal on Sel.
  Areas in Commun.}, vol.~39, no.~5, pp. 1183--1210, 2021.

\bibitem{VoI_USSR}
R.~L. {Stratonovich}, ``On the value of information,'' \emph{Izv. USSR Acad.
  Sci. Tech. Cybern.}, no.~5, 1965.

\bibitem{VoI}
R.~A. {Howard}, ``Information value theory,'' \emph{IEEE Trans. on Systems
  Science and Cybernetics}, vol.~2, no.~1, 1966.

\bibitem{TSC1}
P.~Mayekar, P.~Parag, and H.~Tyagi, ``Optimal source codes for timely
  updates,'' \emph{IEEE Transactions on Information Theory}, vol.~66, no.~6,
  pp. 3714--3731, 2020.

\bibitem{TSC2}
J.~Zhong and R.~D. Yates, ``Timeliness in lossless block coding,'' in
  \emph{Data Compression Conference}, 2016, pp. 339--348.

\bibitem{TSC3}
J.~Zhong, R.~D. Yates, and E.~Soljanin, ``Timely lossless source coding for
  randomly arriving symbols,'' in \emph{IEEE Inf. Theory Workshop}, 2018.

\bibitem{bastopcu2020optimal}
M.~Bastopcu, B.~Buyukates, and S.~Ulukus, ``Optimal selective encoding for
  timely updates,'' in \emph{Annual Conference on Information Sciences and
  Systems}, 2020, pp. 1--6.

\bibitem{Campbell}
L.~L. {Campbell}, ``A coding theorem and {Renyi's} entropy,'' \emph{Information
  and Control}, vol.~8, no.~4, pp. 423--429, 1965.

\bibitem{Larmore}
L.~L. {Larmore}, ``Minimum delay codes,'' \emph{{SIAM} Journal on Computing},
  vol.~18, no.~1, pp. 82--94, 1989.

\bibitem{pappas2021goal}
N.~Pappas and M.~Kountouris, ``Goal-oriented communication for real-time
  tracking in autonomous systems,'' in \emph{{IEEE Intern. Conf. on Autonomous
  Systems (ICAS)}}, 2021.

\bibitem{baer2006source}
M.~B. Baer, ``Source coding for quasiarithmetic penalties,'' \emph{IEEE
  transactions on information theory}, vol.~52, no.~10, pp. 4380--4393, 2006.

\bibitem{cover1999elements}
T.~M. Cover, \emph{Elements of Information Theory}.\hskip 1em plus 0.5em minus
  0.4em\relax J. Wiley \& Sons, 1999.

\bibitem{kosta2020cost}
A.~Kosta, N.~Pappas, A.~Ephremides, and V.~Angelakis, ``The cost of delay in
  status updates and their value: Non-linear ageing,'' \emph{IEEE Transactions
  on Communications}, vol.~68, no.~8, pp. 4905--4918, 2020.

\end{thebibliography}

\end{document}